\documentclass{optica-article}

\journal{opticajournal} 

\articletype{Research Article}
\usepackage{gensymb}
\usepackage{lineno}
\usepackage{float}
\usepackage{amsmath}
\usepackage{xcolor} 
\newcommand{\GKS}[1]{\textcolor{black}{#1}}
\newcommand{\CK}[1]{\textcolor{black}{#1}}

\begin{document}

\title{
Classical-to-quantum transfer of geometric phase for non-interferometric phase measurement and manipulation of quantum state}

\author{Vimlesh Kumar,\authormark{1, \dag} Chahat Kaushik,\authormark{1,2,*,\dag} M. Ebrahim-Zadeh, \authormark{3,4} C. M. Chandrashekar, \authormark{5,6} and G.K. Samanta\authormark{1}}

\address{\authormark{1}Photonic Sciences Lab., Physical Research Laboratory, Ahmedabad, Gujarat  380009, India\\
\authormark{2}Indian Institute of Technology Gandhinagar, Ahmedabad, Gujarat 382424, India\\
\authormark{3}ICFO-Institut de Ciencies Fotoniques, 08860 Castelldefels (Barcelona), Spain\\
\authormark{4}Institucio Catalana de Recercai Estudis Avancats (ICREA), Passeig Lluis Companys 23, Barcelona 08010, Spain\\
\authormark{5}Quantum Optics $\&$ Quantum Information, Department of Instrumentation and Applied Physics, Indian Institute of Science, Bengaluru 560012, India\\ 
\authormark{6}The Institute of Mathematical Sciences, C. I. T. Campus, Taramani, Chennai 600113, India\\
\authormark{\dag}Authors of equal contributions\\
}
\email{\authormark{*}kaushik.28chahat@gmail.com}



\begin{abstract*} 
The geometric phase, originating from the cyclic evolution of a state, such as polarization on the Poincaré sphere, is typically measured through interferometric approaches that often include unwanted contributions from the dynamic phase. Here, we present a non-interferometric technique based on quantum correlation of pair photons to measure the geometric phase of a classical beam. The transfer of geometric phase of the classical pump beam arising from the cyclic evolution of its polarization state on the Poincaré sphere onto the polarization-entangled pair photons generated via spontaneous parametric down-conversion in a Sagnac interferometer enables easy control over the quantum state. Characterization of the generated quantum states reveals that the geometric phase of the pump beam controls the coincidence counts, entanglement visibility, Bell's parameter, quantum state tomography, and fidelity in close agreement with theoretical predictions. We observe sinusoidal modulation of the Bell's parameter and state fidelity with changes in the geometric phase, resulting in transitions between orthogonal Bell states and Bell-like maximally entangled states. Our results establish the geometric phase of the classical pump as a tunable parameter for quantum state control, offering a compact, passive platform for phase manipulation in quantum photonic systems, enabling geometric phase-based quantum gates, and compensating unwanted phase acquired by the quantum state on propagation.
\end{abstract*}

\section{Introduction}

The geometric phase, also known as the Pancharatnam–Berry phase \cite{Pancharatnam:56, Berry:87}, arises from the cyclic evolution of a classical or quantum state on its parameter space, has found widespread applications across classical and quantum optics, including quantitative phase imaging \cite{Bouchal:19}, clock interferometry for enhanced metrology \cite{Zhou:25}, optimal output coupling in oscillators \cite{Chahat:2023}, frequency shifting \cite{Simon:88}, and quantum gates for quantum computing and quantum key distribution \cite{Sjoqvist:08}. In optics and photonics, the geometric phase is readily implemented through the evolution of the polarization state of the light beam on the Poincaré sphere, typically by propagation through various combinations of wave plates sequentially \cite{Simon:89}. Such a simple realization of the dispersion-free phase has enabled accessing robust, broadband elements and systems for beam-shaping \cite{Conrads:25}, atomic optics \cite{Decamps:2017}, generation of spatially structured beams \cite{Jisha:21}, and advanced interferometry and sensing \cite{Garoi:22, Aleman-Castaneda:19}.

Despite significant advances, conventional interferometric methods for measuring the geometric phase \cite{Chyba:88, Ericsson:05} remain challenging due to unavoidable contributions from the dynamic phase. Efforts have been made to mitigate this issue using stable Sagnac interferometers \cite{Chahat:2023}, which reduce the need for active optical path-length stabilization. More recently, non-interferometric approaches have gained attention, including the use of anisotropic lenses composed of multiple spatial light modulators \cite{Malhotra:18}, polarization-dependent diffraction and polarimetry \cite{Arteaga:20}, and nonlinear parametric processes such as stimulated second harmonic generation \cite{Chahat:24}. Nevertheless, there remains a need for novel, non-interferometric techniques that offer simpler experimental implementation and straightforward estimation of the geometric phase in both classical and quantum systems.

On the other hand, imprinting the geometric phase to quantum states through the evolution of polarization on the Poincaré sphere has gained momentum. Notable examples include the generation of energy-time entanglement \cite{Jha:08}, realization of universal quantum gates \cite{Sengupta:24} by imprinting geometric phase onto pair photons generated through spontaneous parametric down-conversion (SPDC) process \cite{Armstrong:62}, and the observation of phase memory in NOON (N = 2) states by transferring either the geometric phase \cite{Qi:20, Huang:24, Kim:06} or dynamic phase \cite{Heuer:14} of the pump beam. However, to date, no reports have demonstrated geometric-phase-induced control of quantum states such as maximally entangled Bell states \cite{EPR:35, Aspect:82, Bell:64}, foundational for numerous quantum applications, including quantum communication \cite{Mishra:22}, quantum teleportation \cite{Bennett:93}, and quantum computing \cite{Sjoqvist:08, Barz:15}.

Here, we report a simple and novel experimental scheme for the non-interferometric measurement of the geometric phase and control of quantum states through the phase memory effect. By transferring the geometric phase of a classical pump beam, induced by rotating a half-wave plate, to polarization-entangled pair photons generated via spontaneous parametric down-conversion (SPDC) in a Sagnac interferometer, we imprint the phase memory of the pump onto the pair photons. The measurement of the geometric phase of a classical beam through the coincidence counts, entanglement visibility, Bell's parameter, quantum state tomography, and fidelity of the quantum state confirms the non-interferometric nature of the method. Using the geometric phase of the pump as a control parameter, we demonstrate state transitions between orthogonal Bell states, typically achieved through Pauli gates, and Bell-like maximally entangled states. The current findings result in a new pathway for realizing phase-dependent quantum operations in a simple and compact experimental setup, and pave the way for future implementation in integrated photonic quantum circuits.

\section{Theory}
We employ the analytical approach to understand the transfer of the geometric phase (GP) of the classical pump beam to the entangled two-photon state and establish the relation of GP to the entanglement state and the Clauser-Horne-Shimony-Holt (CHSH) form of Bell's inequality \cite{CHSH:69} ($S$). To introduce GP into the classical pump beam, we have used the Jones matrix \cite{Jones:41} formalism for the QHQ setup consists of a quarter-wave plate (QWP, Q), a half-wave plate (HWP, H), and another quarter-wave plate (QWP, Q), sequentially positioned with their fast axes oriented at angles $45\degree$,  $\theta_H$, and $45\degree$, with respect to the vertical, respectively. The transfer matrix for the QHQ can be written as \cite{Chahat:2023} 

\begin{equation}\label{1}
    T= QHQ= \begin{pmatrix} e^{2i\theta_H} & 0\\0 & -e^{-2i\theta_H}\end{pmatrix}
\end{equation}

Now, for the classical pump beam of diagonal polarization, $|D\rangle$ = 1/$\sqrt{2}$($|H\rangle$ + $|V\rangle$), (a vector sum of horizontal and vertical polarizations), on propagation through the QHQ setup transform into the polarization state given as 1/$\sqrt{2}$($|H\rangle + |V\rangle$) $\rightarrow$ 1/$\sqrt{2}$($e^{2i\theta_H} |H\rangle-e^{-2i\theta_H} |V\rangle$). The $|H\rangle$ and $|V\rangle$ polarization states acquire geometric phases of equal magnitude, $\phi$= 2$\theta_H$, but with opposite signs, dictated by the solid angle subtended by their cyclic evolution paths on the Poincaré sphere in opposite directions. The polarization projection of the beam 
for diagonal/anti-diagonal and horizontal/vertical basis 
results in laser power of \GKS{$P_{DA}$ = (sin$^22\theta_H$$|D\rangle$ + cos$^22\theta_H$ $|A\rangle$)} and $P_{HV}$ = 1/2($|H\rangle$ + $|V\rangle$), respectively. Therefore, the measurement of the laser at one of the output ports of PBS will result in the power variation for diagonal/anti-diagonal basis and constant power in horizontal/vertical basis for the rotation of half-wave plate (H) of QHQ setup; thus, the geometric phase.

Now the generation of entangled photons using the pump beam of polarizations state,  1/$\sqrt{2}$($e^{i\phi} |H\rangle$ - $e^{-i\phi} |V\rangle$) from a type-0 phase-matched crystal in a the polarization-based Sagnac interferometer \cite{Jabir:17, singh2023}, results in the entangled state given as,

\begin{equation}
\label{eq:2}
\begin{split}
   |\psi_{\phi}^-\rangle\rightarrow \frac{1}{\sqrt{2}}(e^{i\phi}|V\rangle_s |V\rangle_i -e^{-i\phi} |H\rangle_s |H\rangle_i) 
   \\
     \equiv \frac{1}{\sqrt{2}}(|H\rangle_s |H\rangle_i -e^{i2\phi} |V\rangle_s |V\rangle_i)
\end{split}
\end{equation}

Here, the subscripts denote the signal ($s$) and idler ($i$) photons. Neglecting the global phase, we find that the output state, as given by Eq. \ref{eq:2}, resembles a Bell state, with a relative phase introduced by the geometric phase between the orthogonal components of the pump beam. If the state $|\psi_\phi^-\rangle$ propagates through a birefringent medium, then the $|H\rangle_s |H\rangle_i$ and $|V\rangle_s |V\rangle_i$ will acquire different phases, let's say $\phi_H$ and $\phi_V$, respectively, and alter the entangled state. In such a case, the resultant state can be 
\begin{equation}
\label{eq:2a}
\begin{split}
   |\psi_{\phi}^-\rangle \equiv \frac{1}{\sqrt{2}}(|H\rangle_s |H\rangle_i -e^{i(2\phi + \phi_V - \phi_H)} |V\rangle_s |V\rangle_i)
\end{split}
\end{equation}
It is evident from Eq. \ref{eq:2a} that by suitably adjusting the geometric phase, $\phi$, one can make the $2\phi + \phi_V - \phi_H = 0$ to compensate dispersion of the birefringent medium and restore the desired quantum state \cite{Kim:06, Kim:19}.

To study the generated state as shown by Eq. \ref{eq:2}, we calculate, using the simple mathematical treatment \cite{Mandel:95, Tunal:22}, the coincidence probability after polarization projection for the signal and idler photon as

\begin{equation}
\label{eq:7}
\begin{aligned}
   P(\Theta_s,\Theta_i)= \frac{1}{2} (Cos^2\Theta_s Cos^2\Theta_i+ Sin^2\Theta_s Sin^2\Theta_i- \frac{1}{2}Sin2\Theta_s Sin2\Theta_i Cos2\phi)
\end{aligned}
\end{equation}
Here, $\Theta_s$ and $\Theta_i$ denote the inclination angles of the polarizers with respect to the x-axis, placed in the signal and idler arms, respectively. Using the Eq.\ref{eq:7},  the probability of projecting the state onto the $H$, $D$, $V$, and 
$A$ bases can be determined by substituting values of $\Theta_s$  as 0$^\circ$, 45$^\circ$, 90$^\circ$, 135$^\circ$ and scanning $\Theta_i$ from 0$^\circ$ to 180$^\circ$. Again, it is evident from Eq. \ref{eq:7}, that the coincidence probability of the state is dependent on the geometric phase $\phi$. For the D/A basis resulting from the combination of ($\Theta_s$, $\Theta_i$) values of ($\pm$45$^{\circ}$, $\pm$45$^{\circ}$), the coincidence probability is given as 
\begin{equation}
\label{eq:8}
\begin{aligned}
   P= \frac{1}{4} (1- Cos2\phi)
\end{aligned}
\end{equation}
It is evident from Eq. \ref{eq:8} that the coincidence probability of the generated state in the D/A measurement basis depends on the geometric phase. On the other hand, in the H/V basis, the coincidence probability (P = 1/2) remains independent of the geometric phase of the input classical pump beam. Therefore, one can measure the coincidence of the pair photons to measure the geometric phase.

To verify the quantum nature, using Eq. \ref{eq:7}, we calculated the CHSH form of Bell's inequality \cite{CHSH:69} of the generated quantum state as,  
\begin{equation}
\label{eq:10}
\begin{split}
   C(\Theta_s,\Theta_i) &= P(+,\Theta_i,+,\Theta_s) + P(-,\Theta_i,-,\Theta_s)- P(+,\Theta_i,-,\Theta_s) -P(-,\Theta_i,+,\Theta_s) \\
   &= [(Cos2\Theta_s Cos2\Theta_i) - Sin2\Theta_s Sin2\Theta_iCos2\phi]
\end{split}
\end{equation}

Here, $C(\Theta_s,\Theta_i)$, the correlation between the two observables identified by the polarizer angles, $\Theta_s$ and  $\Theta_i$, is averaged over the ensemble of all possible outcomes. Also, the outcome +/- is identified as the emergence and no emergence of photons from the polarizer, respectively. Using Eq. \ref{eq:10}, we find the CHSH Bell's parameter, S, associated with the angles ($\Theta_s$, $\Theta_i$, $\Theta'_s$, $\Theta'_i$) = (0$\degree$, 22.5$\degree$, 45$\degree$, 67.5$\degree$), as

\begin{equation}
\label{eq:12}
\begin{split}
   S &=|C(\Theta_s, \Theta_i) - C(\Theta_s, \Theta_i')|+|C(\Theta_s', \Theta_i) + C(\Theta_s', \Theta_i')|\\
   &= |\sqrt{2}|+|\sqrt{2}Cos2\phi|
\end{split}
\end{equation}
It is evident from Eq. \ref{eq:12} that CHSH Bell's parameter\cite{CILDIROGLU2024,CILDIROGLU2021}, S, of the output pair photons state (given by Eq. \ref{eq:2})  depends on the geometric phase of the parent classical pump beam. It is also interesting to note that the CHSH Bell's parameter has a maximum value of $2\sqrt{2}$ for GP of $\phi$ = 0$^\circ$ and a minimum of $\sqrt{2}$ for $\phi$ = 45$^\circ$. \textcolor{black}{On the other hand, using the von Neumann entropy of the reduced density matrix approach~\cite{Nielsen:91, Bennett:96}, we calculated the entanglement entropy of subsystem $A$ in the bipartite system $A \cup B$, given by $S_A = -\mathrm{Tr} \left( \rho_A \log \rho_A \right)$, where $\rho_A$ is obtained by tracing out subsystem $B$. The two-qubit state represented by Eq.~\ref{eq:2} results in an entanglement entropy of unity, independent of the geometric phase, confirming that the quantum state is maximally entangled. On the other hand, the observed decrease in the CHSH Bell's value with the geometric phase indicates a deviation of the generated quantum state from the standard Bell state, even though the entanglement entropy remains unchanged. To achieve maximal violation of the CHSH Bell's inequality for arbitrary values of $\phi$, it is necessary to determine the optimal measurement bases. Thus, the combined variation of the CHSH Bell's value and entanglement entropy captures the quantum nature of the geometric phase.}

Further, we can calculate the fidelity of the geometric phase induced quantum state, $|\psi_\phi\rangle$, with the initial state, $|\psi_{\phi=0}^-\rangle$, as
\begin{equation}
\label{eq:13}
\begin{aligned}
   F = |\langle\psi_\phi|\psi_{\phi=0}^-\rangle|^2 = Cos^2\phi
\end{aligned}
\end{equation}

The Eq. \ref{eq:13} reveals that the quantum state fidelity depends on the geometric phase of the classical pump beam. As the fidelity $F$ = 0 for $\phi$ = 90$\degree$, we observe the generation of orthogonal states for $\phi$ = 0 and 90$\degree$. Thus, by tuning the geometric phase of the classical pump beam, we can make the transition between two orthogonal Bell states. Furthermore, by measuring the fidelity of the output state, we can also determine the geometric phase. This derivation highlights the possibility of transferring geometric phase from a classical system to a nonclassical state and subsequently measuring it through quantum parameters using any of the equations, Eq. \ref{eq:8}, \ref{eq:12}, and \ref{eq:13}, without using any interferometer commonly used for geometric phase measurement. Moreover, the geometric phase of the pump beam, which can be manipulated by simply rotating a single wave plate of QHQ combination, can be adjusted to compensate for the relative phase acquired by the entangled state during propagation through different media. 

\section{Experimental methods}

\begin{figure}[ht]
\centering
\includegraphics[width=\linewidth]{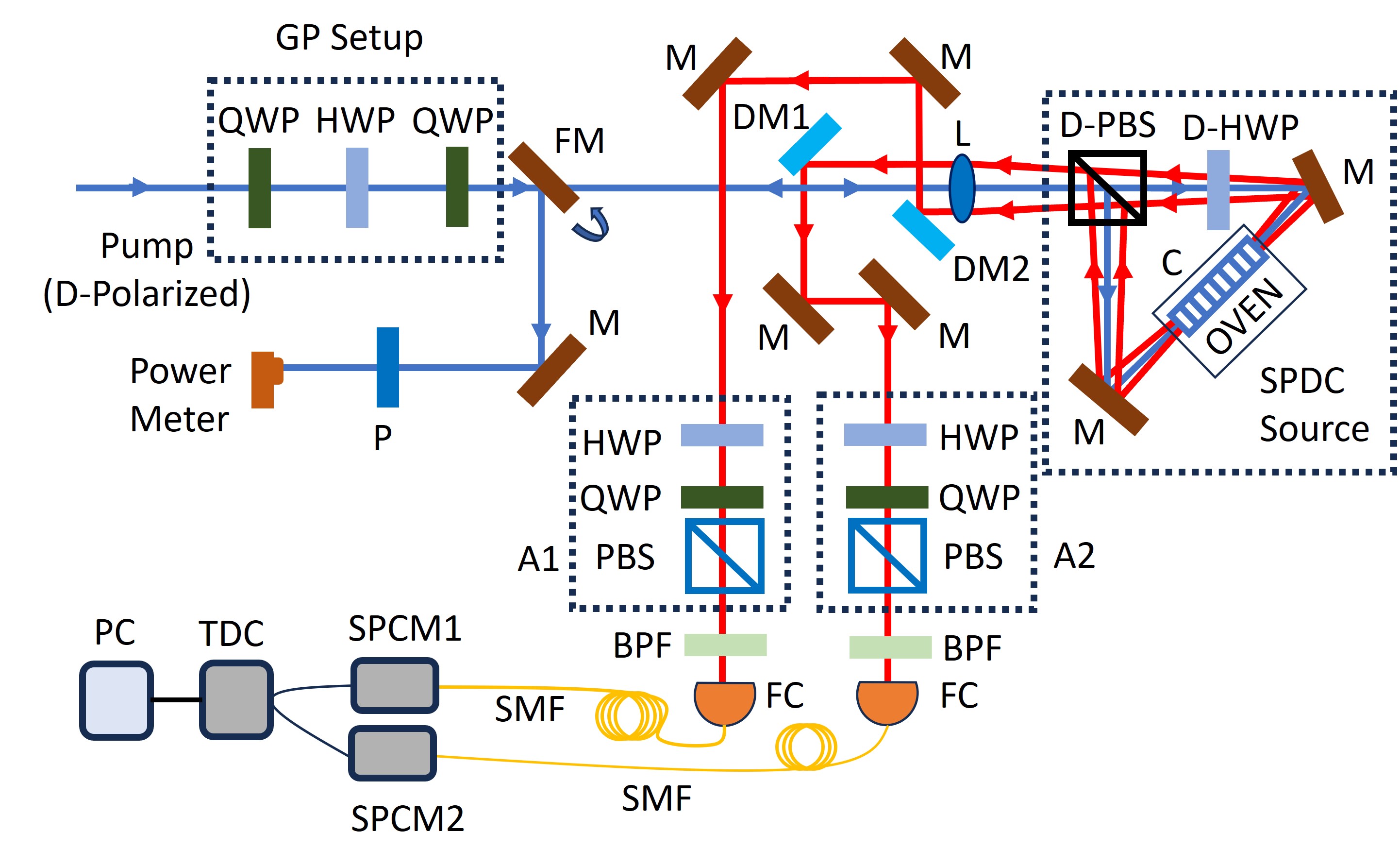}
\caption{Schematic experimental setup for studying the classical geometric phase using quantum states. Blue line denotes the pump at 405 nm, and red line denotes SPDC photons at 810 nm; QWP, Q: quarter wave plate; HWP, H: half wave-plate; GP: geometric phase; FM: flip mirror; M: mirror; P: polarizer, combination of HWP and polarizing beams splitter at 405 nm; D-PBS: dual wavelength (405 nm and 810 nm) polarizing beam-splitter; D-HWP: dual wavelength (405 nm and 810 nm) half-wave plate; C: PPKTP crystal; DM1-2: D-shaped mirror; A1-2: polarization analyzers consist of HWP, QWP, and PBS; FC: Fiber collimator; SMF: single mode fibre; BPF: bandpass filter; SPCM: single photon counting module; TDC: time to digital converter.}
\label{Figure 1}
\end{figure}

The schematic of the experimental setup is shown in Fig. \ref{Figure 1}. A single-frequency, linearly polarized (vertically polarized) continuous-wave (cw) laser (Toptica, TopMode) \cite{Jabir:17, VKumar22}, providing 40 mW of single-frequency output at 405 nm, is used as the pump laser in the experiment. A half-wave plate (not shown in Fig. 
\ref{Figure 1}) transforms the laser output into diagonal (D) polarization. The pump beam then propagates through the geometric phase setup (GP setup), which consists of a quarter-wave plate (QWP, Q), a half-wave plate (HWP, H), and another quarter-wave plate (QWP, Q), arranged in a QHQ configuration with their optic axes oriented at 45$^\degree$, $\theta_H$, and 45$^\degree$ with respect to the vertical, respectively. As a result, the horizontal and vertical components of the pump beam acquire geometric phases of the same magnitude but with opposite signs.
%
%
The geometric phase is imprinted on the pump beam by simply rotating the HWP of the GP setup and verified by using a flip mirror (FM) on the beam path to direct the pump beam for polarization projection using the polarizer (P), a combination of HWP and polarization beam splitter (PBS) optimized for 405 nm,  and the power meter. A lens (L) of focal length $f$ = 150 mm is used to focus the pump at the center of a 30-mm long and 2 x 1 aperture, periodically poled PPKTP crystal (C) of a single grating period of 3.425 $\mu$m. The crystal is mounted on the temperature oven to maintain the crystal temperature of 27$^\circ$ C with a stability of $\pm$0.1$\degree$ C for type-0 ($e\rightarrow e + e$), non-collinear phase-matched, degenerate SPDC photons at 810 nm. The crystal, C, is placed inside the polarization Sagnac interferometer, comprised of a dual-wavelength (405 nm and 810 nm) polarization beam splitter cube (D-PBS), dual-wavelength half-wave plate (D-HWP), and mirrors M, such that the center of the crystal is at equal optical distance from D-PBS for clockwise (CW) and counterclockwise (CCW) beams. The working principle for the generation of entangled photons using the current experimental scheme can be found in Ref. \cite{Jabir:17, singh2023}. The SPDC photons generated in CW and CCW pump beams have an annular ring intensity profile (as observed using an EMCCD) with mutually orthogonal polarization. The SPDC rings generated by the CW and CCW pump beams in the Sagnac interferometer are superimposed by the D-PBS, collimated using the lens (L), and then split into two halves by the D-shaped mirrors (DM1, DM2). These halves are subsequently guided by the mirrors (M) towards the polarization analyzers, A1 and A2, comprised of the half-wave plate (HWP), quarter-wave plate (QWP), and polarization beam splitter (PBS). The transmitted photons from each PBS are spectrally filtered using a band-pass filter (BPF) with a full-width at half maximum (FWHM) of $\sim$3 nm, centered at 810 nm. The filtered photons are then collected into a single-mode fiber (SMF) via a fiber collimator (FC) and detected using a single-photon counting module (SPCM). Finally, the coincidence detection of photons is performed using a time-to-digital converter (TDC) connected to the SPCMs. Throughout the manuscript, the coincidence window is set to 1.6 ns unless stated otherwise.


\section{Results and discussions}

\begin{figure}[ht]
\centering
\includegraphics[width=1\linewidth]{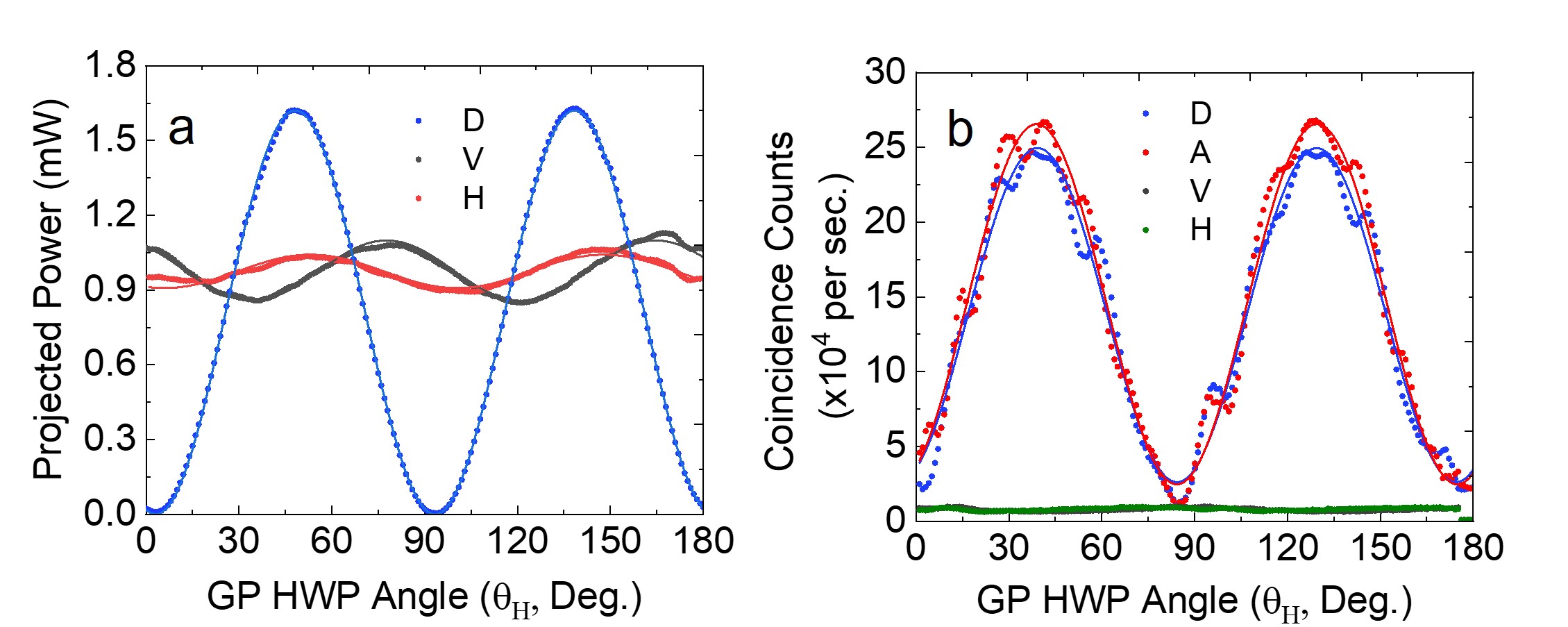}
\caption{ \textbf{(a)} Pump power measurement with GP HWP rotation for polarizer set at 0$^\circ$, 45$^\circ$, and 90$^\circ$ for H (\CK{red}), D (\CK{blue}), and V (\CK{black}) basis respectively; \textbf{(b)} Coincidence Counts measurement with GP HWP rotation for H-basis (\CK{green}), V-Basis (\CK{black}), D-Basis (\CK{blue}), A-Basis (\CK{red}) respectively. }
\label{Figure 2}
\end{figure}

Firstly, we verified the presence of a geometric phase in the classical pump beam. To achieve this, we set the input beam polarization to GP setup (see Fig. \ref{Figure 1}) as diagonal and used the flip mirror (FM) to direct the beam towards the polarization projection unit (P), which consists of a half-wave plate (H) and a PBS (not shown in Fig.). By adjusting the angle of the half-wave plate, we set different measurement bases and recorded the power of the horizontally polarized component transmitted through the PBS. The experimental results are presented in Fig. \ref{Figure 2}(a). As evident from Fig. \ref{Figure 2}(a), the power of the input beam for a diagonal projection varies sinusoidally from a minimum of 0.004 mW to a maximum of 1.63 mW with the geometric phase, which arises from the orientation of the HWP angle, $\theta_H$ in the GP setup. In contrast, for a horizontal/vertical basis, the output power remains constant at half of the total input power. This observation aligns with theoretical expectations, where the PBS output for diagonal/anti-diagonal and horizontal/vertical bases is given by $P_{D/A}$ = sin$^2\phi$ and $P_{H/V}$ = 1/2, respectively. These results confirm that the geometric phase is imprinted on the input classical pump beam simply by rotating the half-wave plate in the GP setup. The small oscillations observed in the output power on the horizontal/vertical measurement basis can be attributed to the unoptimized retardance of the waveplates and the related compensation through waveplate tilt used in the experiment.

Further, we studied the effect of the geometric phase of the input pump on the pair photons. In doing so, we removed the flip mirror (FM) from the pump beam path and detected the pair photons of diametrically opposite points of the SPDC ring for different polarization projections using the different settings of the analyzers, A1 and A2. We set analyzer A1 at the polarization settings of H, V, D, and A and the corresponding orthogonal polarization setting in analyzer A2 for minimum coincidence counts and rotated the HWP of the GP setup with the results shown in Fig. \ref{Figure 2}(b). As evident from Fig. \ref{Figure 2}(b), the coincidence counts of the pair photons show the effect of the geometric phase on the D/A basis in agreement with Eq. \ref{eq:8}, whereas there is no effect in the H/V basis exactly the same as the pump beam (see Fig. \ref{Figure 2}(a)). This observation not only confirms the direct transfer of the geometric phase from the classical pump beam to the single photons but also demonstrates a novel approach to measuring the geometric phase of a classical pump beam. Analyzing the coincidence counts of the generated pair photons provides a simple and effective alternative to conventional interferometric setups typically used for phase measurement. Notably, this technique is inherently free from dynamic phase contributions, making it a robust and direct approach for dynamic phase-free geometric phase measurement in quantum optics. 

\begin{figure}[ht]
\centering
\includegraphics[width=1\linewidth]{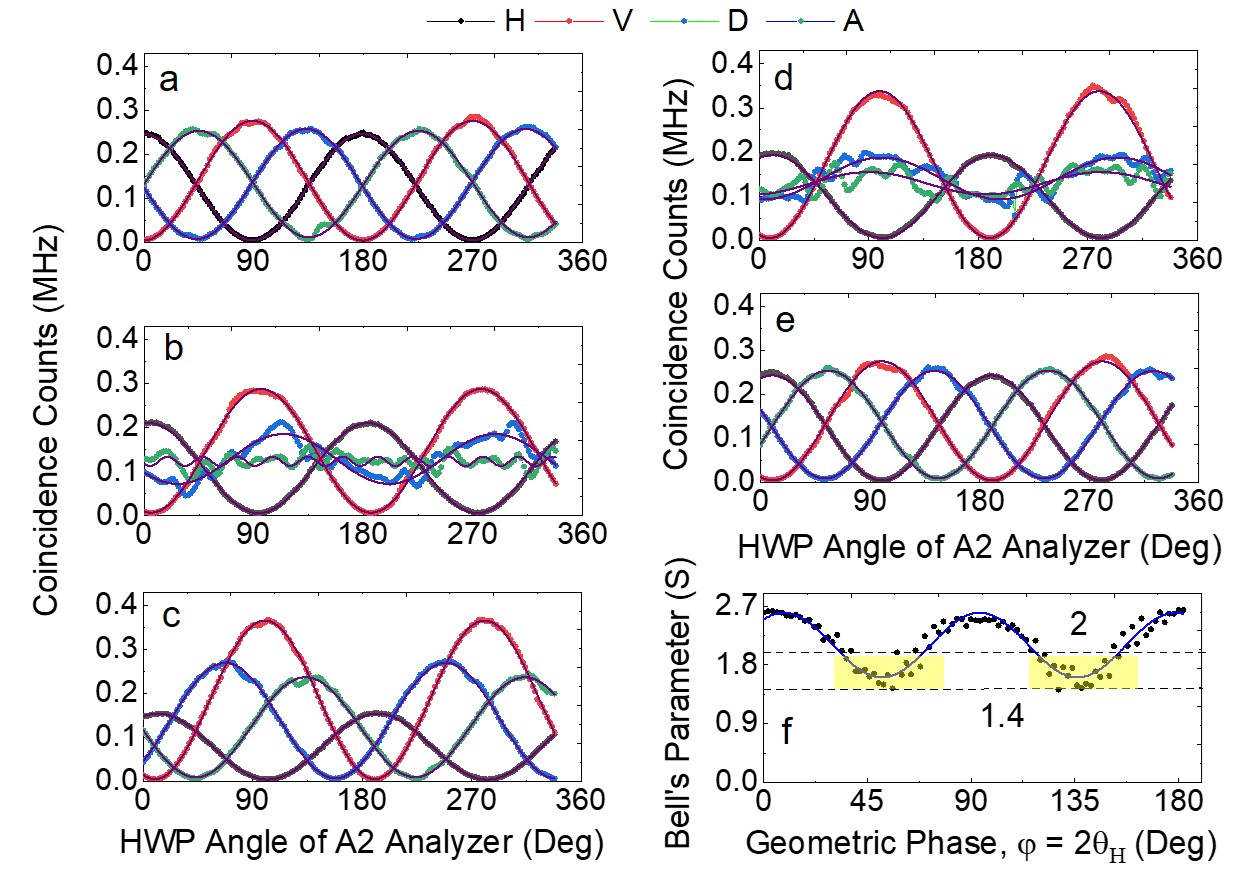}
\caption{Variation of the coincidence counts as a function of the angle of HWP of analyzer, A2, for the angles 0$^\circ$ (H-Basis, \CK{black}),  22.5$^\circ$ (D-Basis, green), 45$^\circ$ (V-Basis, \CK{red}), and 67.5$^\circ$ (A-Basis, \CK{blue}) of HWP of analyzer, A1, for HWP angle, $\theta_H$, of GP setup at (a) 0$^\circ$, (b) 22.5$^\circ$, (c) 45$^\circ$, (d) 67.5$^\circ$, (e) 90$^\circ$. (f) Variation of CHSH Bell's parameter as a function of geometric phase. Yellow shaded regions identify the cessation of entanglement. Solid lines are fits to the experimental data points. }
\label{Figure 3}
\end{figure}
To confirm the transfer of the geometric phase of the classical pump beam to the entangled pair photons, we investigated its effect on the generated quantum state. Keeping the pump power constant at approximately $\sim$2 mW, we studied the quantum interference of the generated state from the Sagnac interferometer-based entangled photon source \cite{Jabir:17, singh2023} (see Fig. ~\ref{Figure 1}). Coincidence counts between the entangled photons were recorded while fixing the angle $\Theta_s$ of the HWP in analyzer, A1, to project onto various polarization bases, and simultaneously rotating the angle $\Theta_i$ of the HWP in analyzer, A2, from 0 – 180$^\circ$, for different values of geometric phase $\phi$ induced by the HWP angle $\theta_H$ of the GP setup. The results are presented in Fig. ~\ref{Figure 3}.

As shown in Fig. ~\ref{Figure 3}(a), for $\phi = 2\theta_H = 0^\circ$, the coincidence counts at four polarization bases, H (black), V (red), D (green), and A (blue) corresponding to $\Theta_s = 0^\circ$, 45$^\circ$, 22.5$^\circ$, and 67.5$^\circ$, respectively, follow the expected sinusoidal modulation. The interference visibility in the H/V and D/A bases was measured to be greater than 0.94 and 0.91, respectively, without applying any corrections to the experimental data. The corresponding measured CHSH Bell parameter $S = 2.63$ confirms the generation of a high-purity entangled state. We repeated this measurement for geometric phases $\phi = 2\theta_H = \pi/4$, $\pi/2$, $3\pi/4$, and $\pi$, with results shown in Fig. ~\ref{Figure 3}(b–e). The interference visibility in the H/V bases remained consistent across all values of geometric phase, with slight variations in maximum coincidence counts attributed to changes in the relative amplitudes of the H and V components of the pump beam (see Fig. \ref{Figure 2}(a)) due to the residual birefringence of the HWP of GP setup. In contrast, while the D/A basis interference visibility remains high and comparable to that in H/V for $\phi = 0$, $\pi/2$, and $\pi$, no interference is observed in D/A for $\phi = \pi/4$ and $3\pi/4$. This can be understood from Eq. ~\ref{eq:2}, representing the generated entangled state in the experiment. Here, the relative phase between $|HH\rangle$ and $|VV\rangle$—originating from the clockwise and counter-clockwise pump beams in the Sagnac loop is $2\phi = 4\theta_H$. As $\theta_H$ varies from $0^\circ$ to $90^\circ$, the state transitions from $|\psi_{\phi}^-\rangle$ to the orthogonal state $|\psi_{\phi}^+\rangle$ at $\theta_H = 45^\circ$, and then returns to $|\psi_{\phi}^-\rangle$. Consequently, interference is preserved in all polarization bases except at $\theta_H = \pi/8$ and $3\pi/8$, where the relative phase $e^{-i4\theta_H}$ renders the two-photon amplitudes distinguishable, thereby suppressing interference in the D and A basis.

To further examine the geometric phase-dependent changes in the quantum state, we measured the CHSH Bell parameter $S$ as a function of geometric phase, with results shown in Fig.~\ref {Figure 3}(f). The experimentally obtained values of $S$ (black dots) follow a sinusoidal variation, and a theoretical fit (black line) reveals the functional form of $S = a + b \cos(4\theta_H)$, closely resembling Eq. ~\ref{eq:12}, with fit parameters $a = \sqrt{2}$ and $b = 1.146 < \sqrt{2}$. The small deviation from the ideal maximum $S = 2\sqrt{2}$ is attributed to experimental imperfections such as multiphoton contributions, imperfect mode coupling, and optical losses. It is worth noting that none of the presented data have been corrected for background, dark counts, or system losses. Inclusion of such corrections would improve the quantitative agreement of the experimental values with theoretical expectations.

\begin{figure}[H]
\centering
\includegraphics[width=\linewidth]{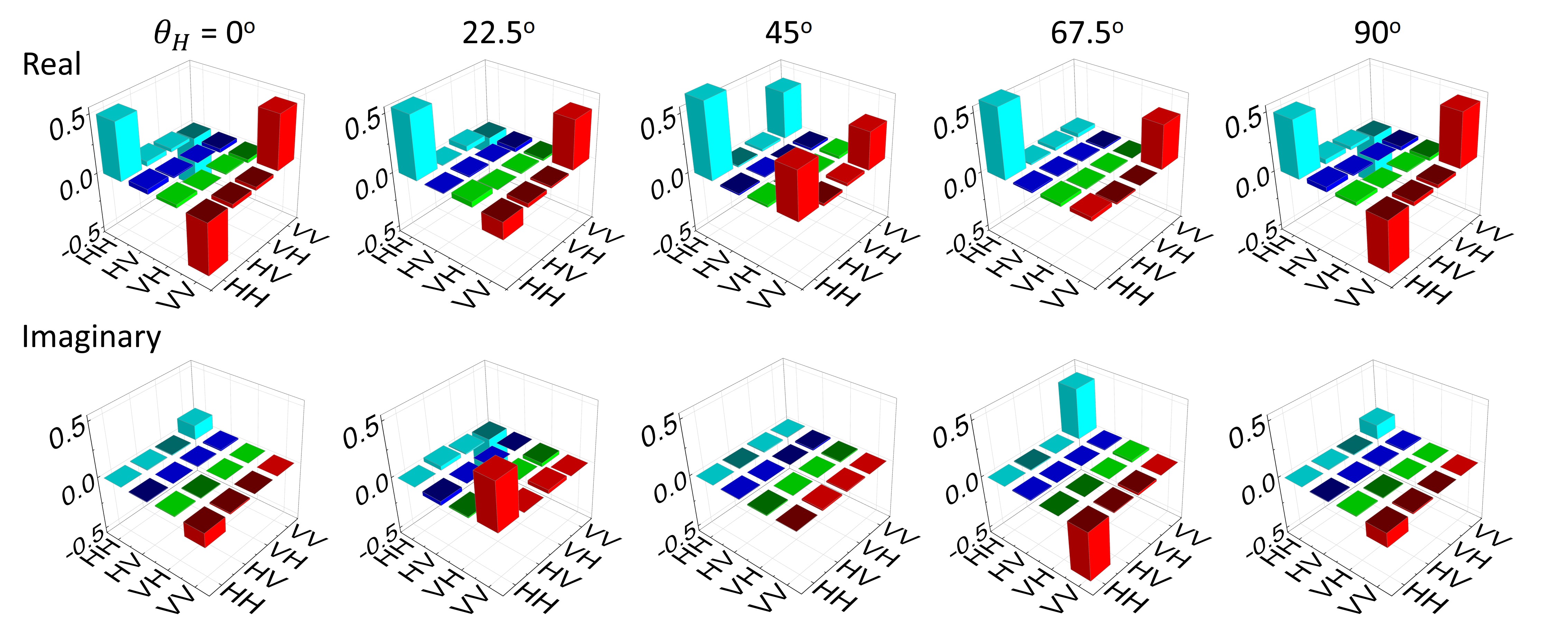}
\caption{Real (top row) and imaginary (bottom row) part of reconstructed density matrix of generated quantum state for different values of geometric phase, $\phi$ of the pump beam arising from HWP angle $\theta_H$ = 0$^\circ$, 22.5$^\circ$, 45$^\circ$, 67.5$^\circ$, and 90$^\circ$.}
\label{Figure 4}
\end{figure}

To gain deeper insight into the generated states, we performed quantum state tomography. Based on coincidence counts recorded for different combinations of polarization projections, we reconstructed the real and imaginary parts of the two-qubit density matrix for various geometric phase values, as shown in Fig. ~\ref {Figure 4}. From the real (top row) and imaginary (bottom row) part of the reconstructed density matrix as shown in Fig. ~\ref{Figure 4} it is evident that the quantum state transitions from $|\psi_{\phi}^-\rangle$ to $|\psi_{\phi}^+\rangle$ and back to $|\psi_{\phi}^-\rangle$ for $\theta_H = 0^\circ$, $45^\circ$, and $90^\circ$, respectively. For intermediate values, $\theta_H = 22.5^\circ$ and $67.5^\circ$, the real part of the density matrix remains largely unchanged, while the sign of the off-diagonal elements in the imaginary part flips, indicating a relative phase shift between $|HH\rangle$ and $|VV\rangle$ components. This corresponds to a quantum state of the form $\frac{1}{\sqrt{2}} \left[ |HH\rangle \mp i |VV\rangle \right]$, in good agreement with the theoretically predicted state in Eq. ~\ref{eq:2}. 

\begin{figure}[ht]
\centering
\includegraphics[width=\linewidth]{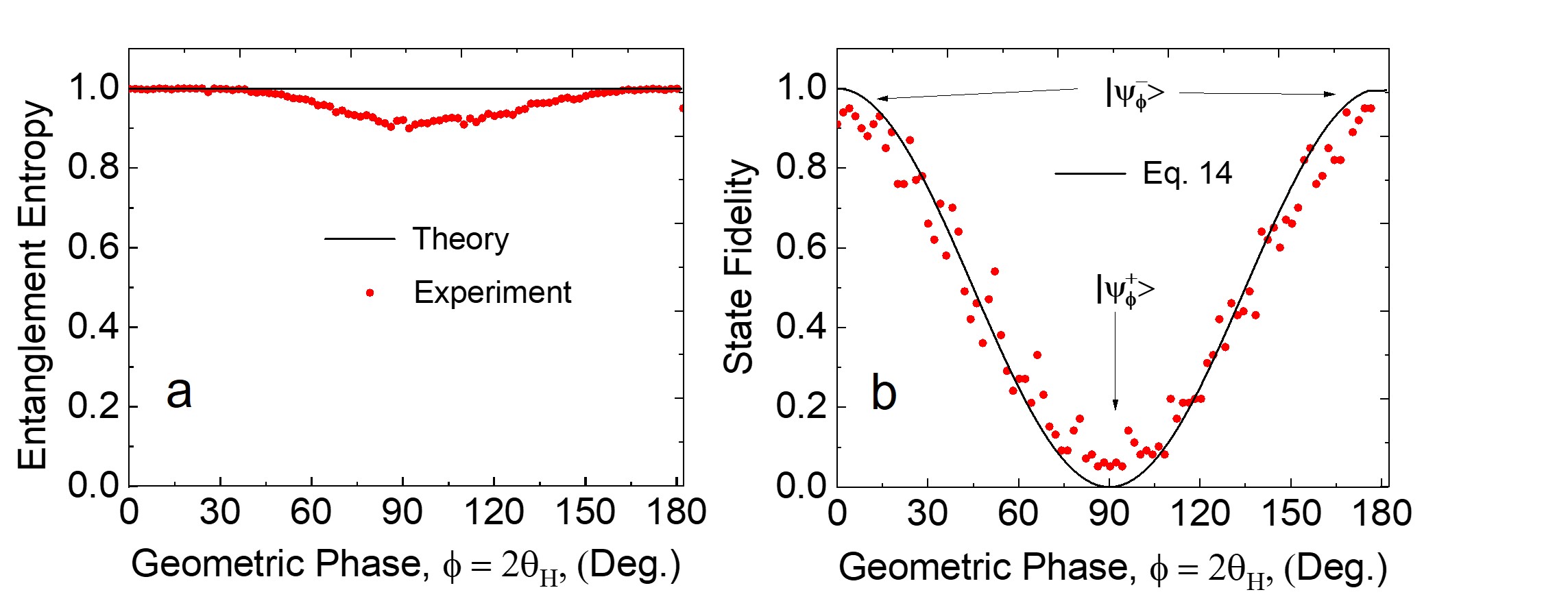}
\caption{ Variation of (a) entanglement entropy and (b) state fidelity of the generated state as a function of the geometric phase of the classical pump beam arising from the HWP angle, $\theta_H$ of the GP setup. Dots and solid lines represent experimental and theoretical results, respectively. }
\label{Figure 5}
\end{figure}

We further investigated the effect of the controlled geometric phase of the classical pump beam on the generated quantum state by analyzing the entanglement entropy and state fidelity, as shown in Fig. ~\ref{Figure 5}. Using the Von Neumann entropy of the reduced density matrix approach ~\cite{Nielsen:91, Bennett:96}, we calculated the entanglement entropy of the experimentally generated two-qubit state as a function of the HWP angle $\theta_H$ in the GP setup. As shown in Fig. ~\ref{Figure 5}(a), the entanglement entropy obtained from the experimentally reconstructed density matrices (red dots) remains nearly constant and close to unity across different values of geometric phase $\phi$, corresponding to various $\theta_H$ values. This agrees well with the theoretical prediction (black dot) calculated from the state given by Eq. ~\ref{eq:2}. A small deviation (approximately 10\%) in the entanglement entropy around $\theta_H = 45^\circ$ can be attributed to the residual birefringence of the HWP in the GP setup, also evident in Fig. \ref{Figure 2}(a), which induces slight power imbalance between orthogonal polarization components of the classical pump beam, thereby affecting the amplitudes of the generated quantum state. These observations confirm that the generated states are maximally entangled. It is also evident that, although the geometric phase of the classical pump beam influences the type of Bell state produced, it does not affect the degree of entanglement.

On the other hand, the state fidelity of the generated geometric-phase-dependent quantum state, calculated with respect to the Bell state $|\psi_{\phi}^-\rangle = \frac{1}{\sqrt{2}}(|HH\rangle - |VV\rangle)$, varies from 0.97 to 0.95 as the HWP angle $\theta_H$ of the GP setup changes from $0^\circ$ to $90^\circ$, reaching a minimum of zero at $\theta_H = 45^\circ$. The experimental result (dots) is in close agreement with Eq. \ref{eq:13}. A small offset observed between the experimental and theoretical data can be attributed to the least count error of the rotational stage used to set the angle of the HWP. Such observation confirms the controlled generation and transition between the states $|\psi_{\phi}^-\rangle$, $|\psi_{\phi}^+\rangle$, and again $|\psi_{\phi}^-\rangle$ by simply tuning the geometric phase of the classical pump beam. Moreover, for intermediate values of $\theta_H$, the generated quantum states deviate from ideal Bell states. Notably, the transformation from $|\psi_{\phi}^-\rangle$ to $|\psi_{\phi}^+\rangle$, which is typically realized by applying single-qubit Pauli operators on one qubit of the quantum state, is achieved here through modulation of the geometric phase of the classical pump beam.

\section{Conclusion}

In conclusion, we have successfully demonstrated the direct transfer of geometric phase arising from the cyclic evolution of the polarization state of a classical pump beam on the Poincaré sphere to a quantum-entangled state. Using quantum interference measurements, we measured the transferred geometric phase, free from dynamic phase contributions, with high precision. Through quantum state tomography, entanglement entropy, Bell inequality tests, and state fidelity measurements, we confirmed that a simple modulation of the geometric phase of the classical pump beam allows for precise generation, control, and manipulation of the quantum state. This includes a controlled transformation between orthogonal Bell states, while preserving maximal entanglement fidelity. We also observe the generation of quantum states distinct from Bell states, exhibiting high entanglement entropy, by imprinting different amounts of geometric phase on the pump beam. The current study further highlights the potential of implementing geometric phase-based quantum gates. The all-optical control method also offers a tunable, stable, compact, and scalable approach for realizing phase-dependent quantum operations, with significant implications for future integrated photonic quantum circuits.

\begin{backmatter}

\bmsection{Acknowledgments} The authors acknowledge support of the Department of Space, Govt. of India. G. K. S. acknowledges the support of the Department of Science and Technology, Govt. of India, through the Technology Development Program (Project DST/TDT/TDP-03/2022)

\bmsection{Data availability} Data underlying the results presented in this paper are not publicly available at this time but may be obtained from the authors upon reasonable request.

\bmsection{Disclosures}
The authors declare no conflicts of interest.

\end{backmatter}
\bibliography{sample}

\end{document}